# Universal Computer Aided Design for Electrical Machines

[1]Aravind Vaithilingam Chockalingam, [2]Ikujuni Grace Olasehinde, [3]Rozita Teymourzadeh
Department of Electrical and Electronic Engineering
Faculty of Engineering, UCSI University
Kuala Lumpur, Malaysia

*Abstract*— **Electrical machines are devices that change either mechanical or electrical energy to the other and also can alternate the voltage levels of an alternating current. The need for electrical machines cannot be overemphasized since they are used in various applications in the world today. Its design is to meet the specifications as stated by the user and this design has to be an economical one. The design therefore revolves around designing the machine to meet the stipulated performance required, the cost available and the lasting life of the machine. This work aims to eliminate the tediousness involved in the manual hand calculations of designing the machines by making use of a graphical user interface and using iterations in situations where the data would have been assumed.**

I. INTRODUCTION

The design of motor in general is a complex procedure and hence requires an easy and understandable design method and tool for the designer. CAD design for this work looks into the design of DC machine, induction machines, synchronous machines, transformers, Rotary SRM and the inductance profile of SRM[1]. The induction motor is divided into the design of three phase and single phase induction motor with each having its own GUI. Synchronous motor is divided into salient and round rotor type and the transformer is divided into the shell and core type. The designs have been made with consideration to the performance of the machine. Other limitations such as cost and insulating material have not being taken into consideration. The design of electrical machines [2], [3] has looked considerable into the design of various machines. The reluctance machine design works on the inductance profile and design of the rotary SRM [1], [4], [5].

II. DESIGN TOOL

The analysis and synthesis way of designing electrical machines have been used in the design of the machines. The synthesis method is used for DC machine, AC machine, and transformers. The analysis method has been used for the design of switched reluctance method [1]. There is a general toolbox interface that introduces the user to the design of each electrical machine. This is as shown in the fig.1

This introductory GUI makes use of radio buttons that links the designer to the various machine design GUI pages.

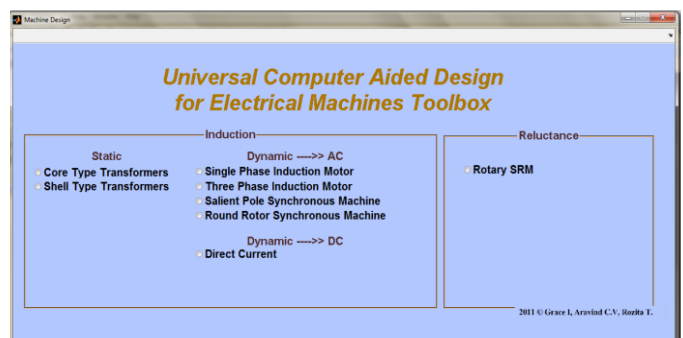

Fig. 1: Introductory GUI

The materials used for the design of electrical machines greatly depend on the hysteresis loop of the machines. Various materials have been used for the design of this toolbox and it is as shown in the fig. 2.

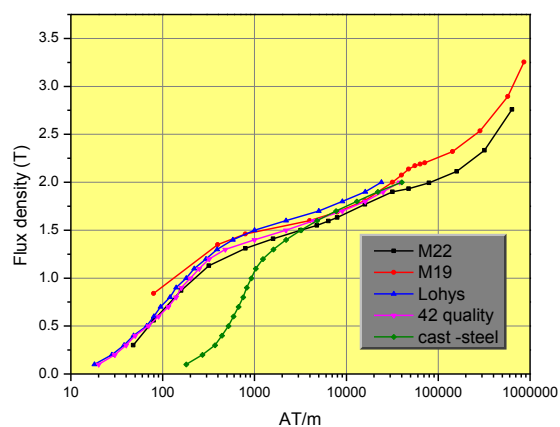

Fig. 2: Magnetization curves

The machine specifications for each machine design GUI include the following:
- The ratings of the machine such as the voltage, KVA, peak current, speed, number of poles.
- For some of the motor design, the required outer surface parameters of the machine are required such as the total diameter and length of the machine.
- For SRM, stator and rotor pole arc, length of air gap required and active or passive element of the linear SRM.

- Number of phases for the machine.

### III. MACHINE DESIGN

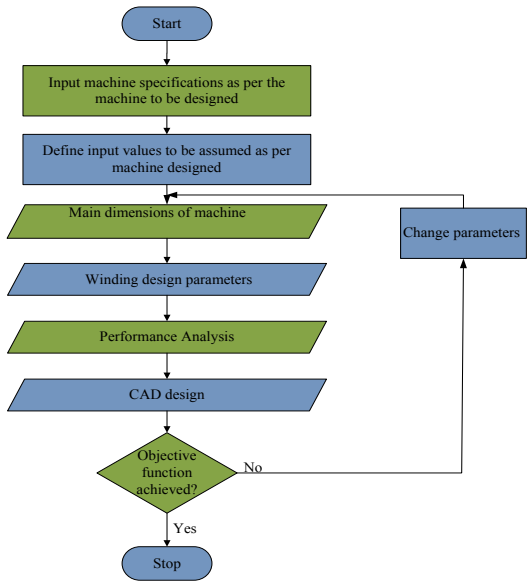

**Fig. 2: Generalized machine design flowchart**

Fig.3 is a generalised design procedure for electrical machines. The analysis on how to design reluctance machine is has been extensively done by [1], [4], [5]. The design of the other electrical machines has being done by [2], [3]. The design procedure looks into the design of the main dimensions, field windings, No load calculations, performance characteristics and others. For all machines the main criteria for the design is getting the output coefficient. The output coefficient is gotten as

$$C_0 = \frac{P}{D^2 \times L \times N} \quad (1)$$

Where P is the output of the machine in KVA, D is the diameter of armature (m) and L is the gross length of armature (m), and N is the speed of the machine in RPM.

Other relevant considerations are that of the magnetic and electric loadings and are specified as follows.

$$P\emptyset = B_{av} \pi DL \text{ , Wb}$$

Where $B_{av}$ = specific magnetic loading (Wb/m$^2$), D is diameter of the armature (m) and L is the length of the armature (m) and P is the number of poles.
The specific magnetic loading,

$$B_{av} = \frac{P\emptyset}{\pi DL} \quad (2)$$

Total electric loading;

$$I_z.Z = \frac{a_c}{\pi D} \text{ , A}$$

Where $I_z.Z$ is the total electric loading, $a_c$ is the specific electric loading in A/m.
The specific electric loading,

$$a_c = \frac{I_z.Z}{\pi D} \quad (3)$$

### IV. RESULTS

#### TRANSFORMERS

Transformers are divided into shell and core type transformers. Although shell type transformers are hardly used, they are still in used in small capacity machines. Both types of transformers have being designed. The GUI enables the user to be able to select the core type, number of phases and the winding required.

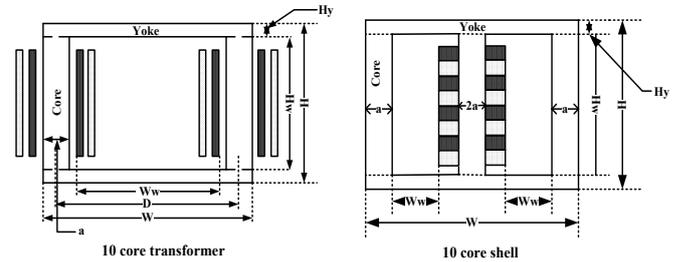

**Fig. 3: Transformers CS**

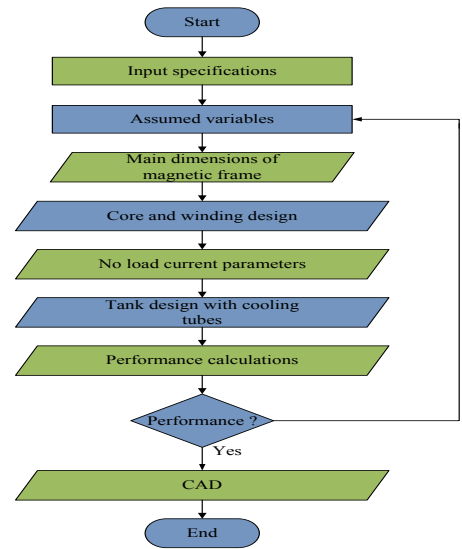

**Fig. 4: Transformer flowchart**

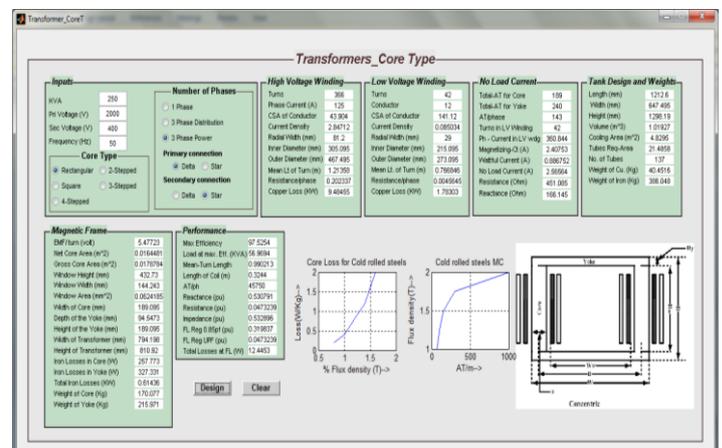

**Fig. 5: Core type GUI result**

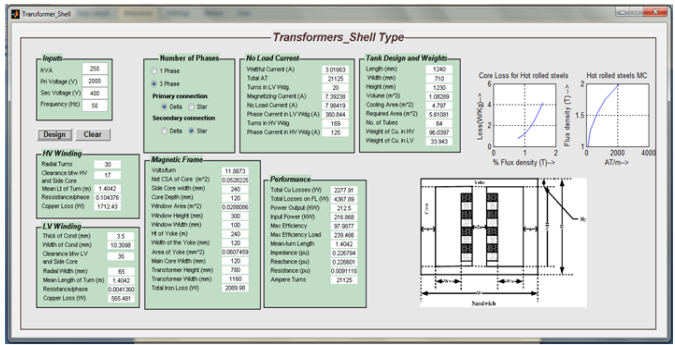

**Fig. 6: shell type GUI result**

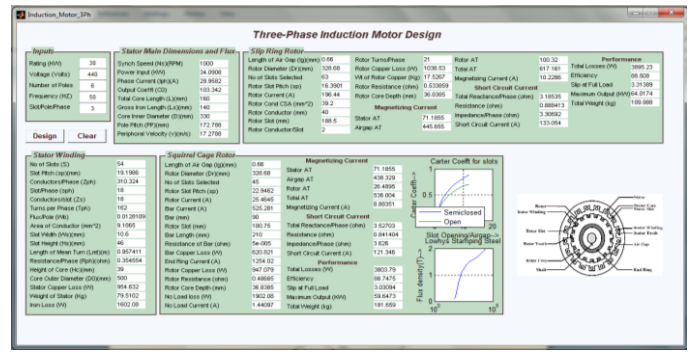

**Fig. 10: Three phase IM GUI result**

For the GUI design, provisions are made so that the designer can pick if the type of core required, number of phases and windings to be used.

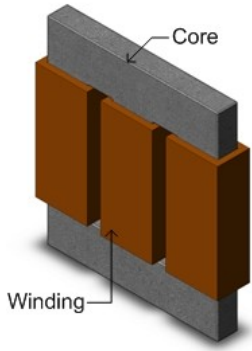

**Fig. 7: Transformer CAD**

INDUCTION MOTORS

The design GUI for the induction motor has been divided into single phase and three phase GUI.

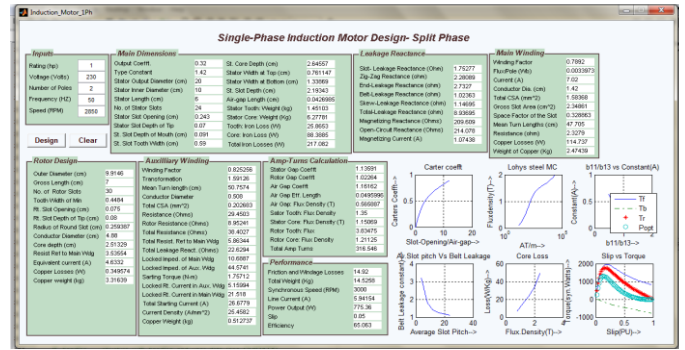

**Fig. 11: Single phase IM GUI result**

The slip of induction motor matters a lot as it greatly affects the torque of the motor. The slip is the difference between the synchronous speed and the actual speed. The need for slip is because an induction motor does not generate voltage when the rotor is running at a synchronous speed to that of the magnetic field. This is the reason why a much difference in the slip, yields a large torque value. The graph as seen from the GUI shows that with increase in the slip of the machine, the torque is greatly increased until it gets to a value where the slow speed in movement of the rotor is not strong enough to cut across the magnetic field thereby resulting in a reduction in the torque produced.

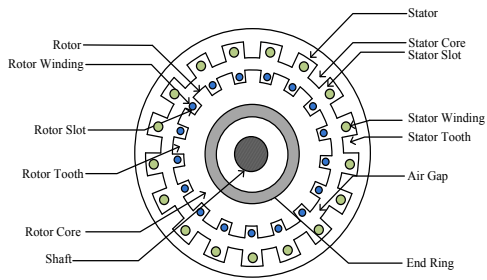

**Fig. 8: Induction motor CS**

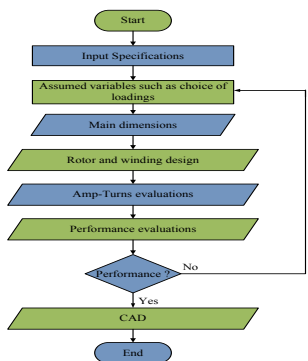

**Fig. 9: IM design flowchart**

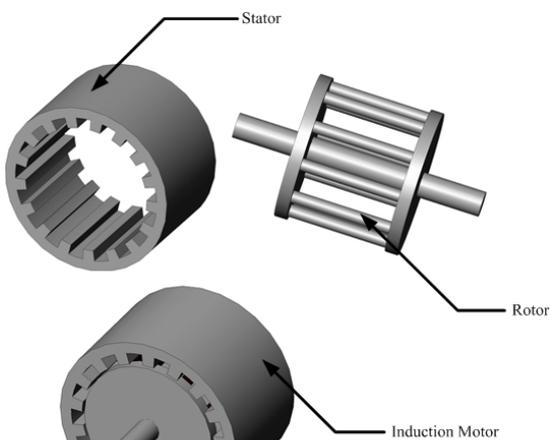

**Fig. 12: Squirrel cage IM CAD**

SYNCHRONOUS MACHINES

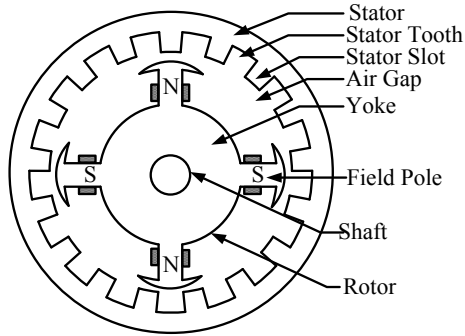

**Fig. 13: Synchronous motor CS**

The synchronous machine GUI shows the plots of the field winding depth, core losses vs. flux density, carter's coefficient for the slots and open circuit characteristics of the machine.

The open circuit characteristics test is taken when the generator is operating without load, running at synchronous speed and the field current is increased from zero to maximum. As seen from the graph in the GUIs', the curve is almost perfectly linear until some saturation is seen at high field currents. This is because the reluctance of the motor at an unsaturated iron frame is far less than that of the air gap thereby all mmf seem to pass through the air gap. As the iron gets saturated, the reluctance increases rather slowly.

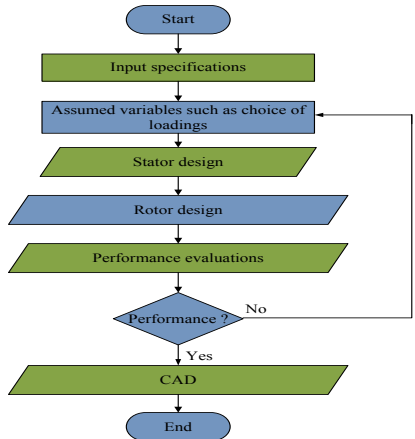

**Fig. 14: Synchronous design flowchart**

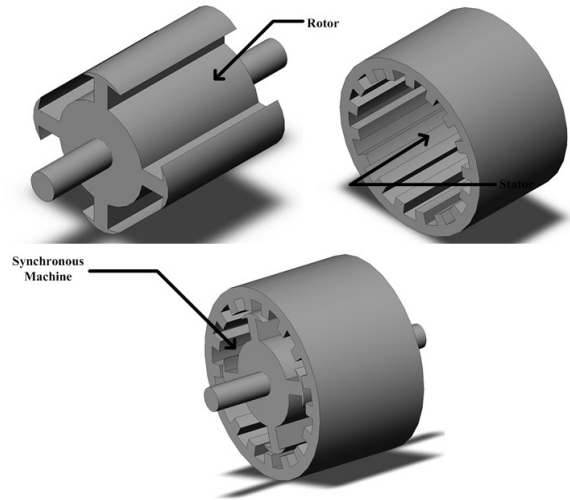

**Fig. 17: Salient pole CAD**

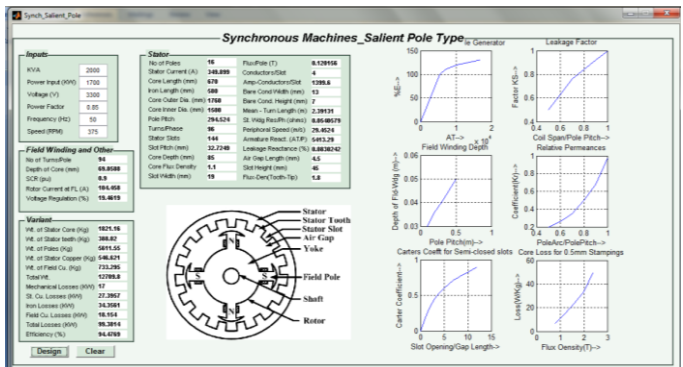

**Fig. 15: Salient pole GUI result**

DC MACHINE

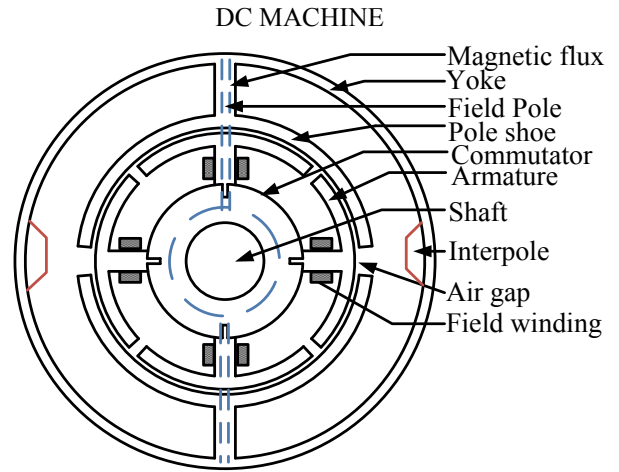

**Fig. 18: DC machine CS**

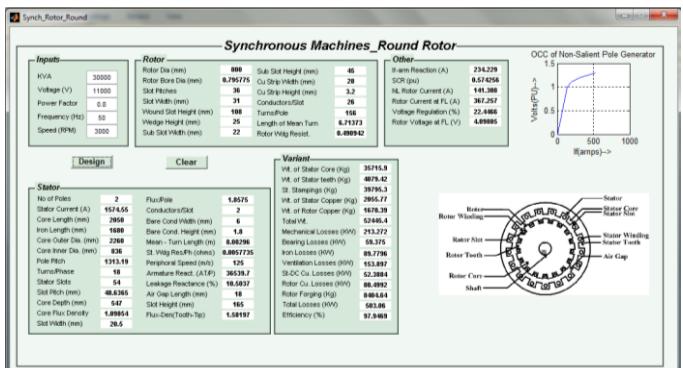

**Fig. 16: Round rotor GUI result**

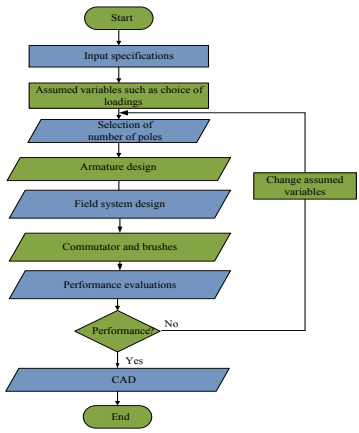

**Fig. 19: DC machine flowchart**

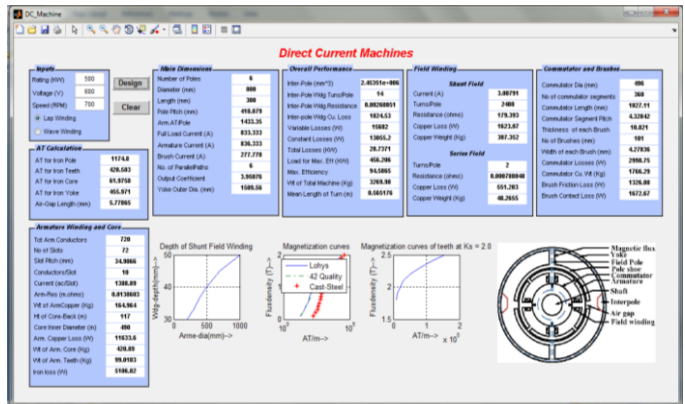

**Fig. 20: DC machine GUI result**

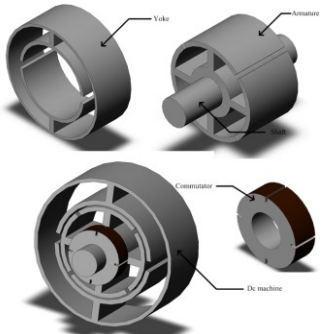

**Fig. 21: DC machine CAD**

SWITCHED RELUCTANCE MOTOR

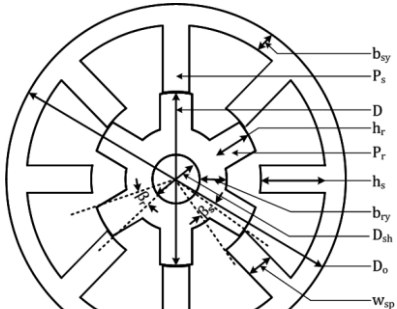

**Fig. 22: SRM CS**

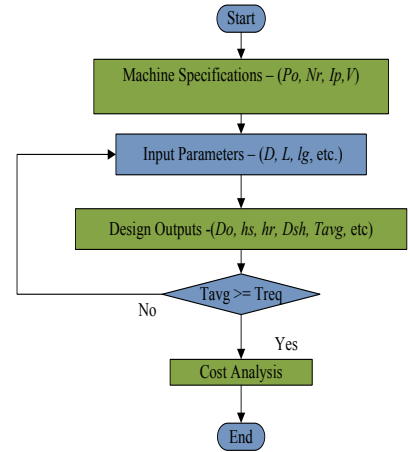

**Fig. 23: SRM design flowchart**

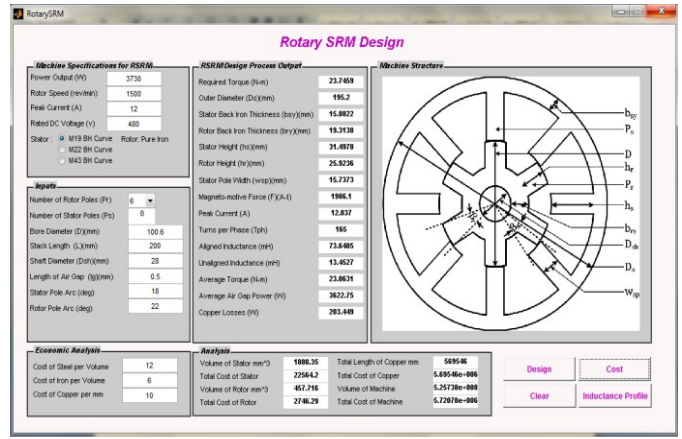

**Fig. 24: Rotary SRM design GUI result**

The SRM design GUI as shown above shows the torque, height of rotor and stator, back iron thickness, pole width, turns per phase, aligned and unaligned inductance and the average air gap power. If the average torque generated does not give the required output, the user is required to change some inputs parameters such as the stator and rotor pole arc. This will help move the average torque closer to the required torque. When this is achieved, the motor can then be said to have been designed.

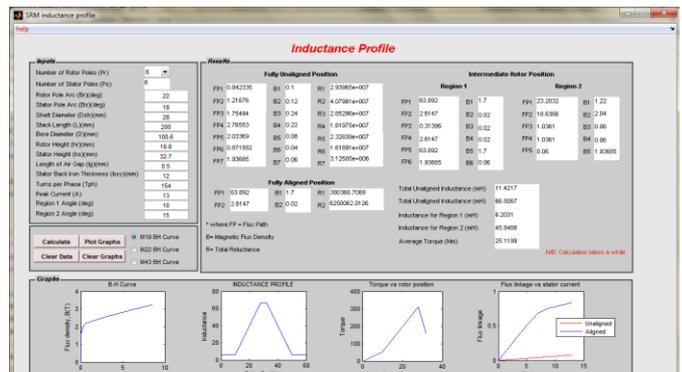

**Fig. 25: SRM inductance profile result**

The inductance profile is a graph of the relationship between the inductance and the rotor position. The inductance profile shows four distinct regions. For the example SRM, the motor is at a fully aligned position when its angle is 30° while 0° is when the motor is at a fully unaligned position. The inductance profile results shown are that of the unaligned inductance, aligned inductance, torque, graphical analysis of some of the performance of the motor and the intermediate position inductance.

The results as shown in the GUI below, shows the inductance for each flux path, the magnetic flux density assumed for that flux path and the reluctance for the flux path. FP in the figure indicates flux path while the number in front indicates the number of flux path. B indicates magnetic flux density assumed for that path and R indicates the reluctance of the motor at that path.

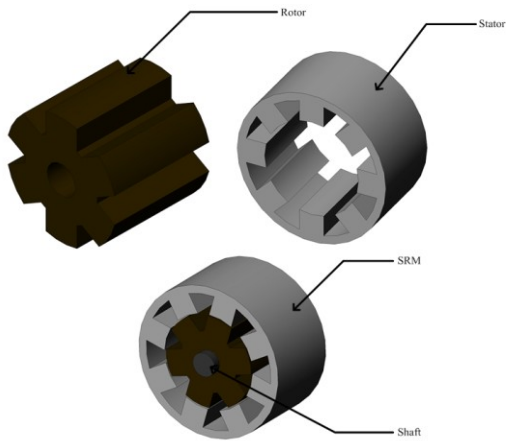

**Fig. 26: SRM CAD**

## V. CONCLUSION

A generalised tool box for optimisation of computer aided design for electrical machines has been built successfully. The SRM design has been verified by using analytical and cyclic integration method. [5], [2]. Others have been verified by using manual calculations. Although these machines have being designed, this may not be the optimum design because it does not put into consideration the cost of the machine, available materials to be used, human labour available, temperature rise and other constraints.

## VI. ACKNOWLEDGMENT

## REFERENCES

[1] Krishnan, R., "Switched Reluctance Motor Drives: Modeling, Simulation, Analysis, Design, and Applications", CRC Press, 2001

[2] Aravind C.V, Kamala K. C. (2003). Design of electrical apparatus. Chennai: Charulatha publications

[3] K.M. Vishnu Murthy (2008). Computer Aided Design of Electrical Machines. Hyderabad: BS Publications

[4] Praveen V., Design of Switched Reluctance Motors and Development of a Universal Controller for Switched Reluctance and Permanent Magnet Brushless DC Motor Drives, 2001, pp. 1-104.

[5] Aravind CV, Samuel Bright., Design of reluctance motors for low speed applications, Bachelor's degree Thesis, May 2011